\documentclass[twocolumn,showpacs,preprintnumbers,amsmath,amssymb]{revtex4-1}

 \usepackage{graphicx}
 \usepackage{amsmath}
 \usepackage{amsfonts}
 \usepackage{amssymb}
 \usepackage{natbib}
 \usepackage{pstricks}
 \usepackage{psfrag}
 \usepackage{epsfig}
 \usepackage[ansinew]{inputenc}

  \newcommand{\eb}[1]{{\color{black} #1}}
  \newcommand{\sr}[1]{{\color{black} #1}}

\begin{document}
\title{Photofocusing: Light and flow of phototactic microswimmer suspension}
\author{Matthieu Martin, Alexandre Barzyk, Eric Bertin, Philippe Peyla, Salima Rafai}
\affiliation{Univ.~Grenoble Alpes, LIPHY, F-38000 Grenoble, France}
\affiliation{CNRS, LIPHY, F-38000 Grenoble, France}

\begin{abstract}
We explore in this \eb{paper} the phenomenon of photofocusing: a coupling between flow vorticity and biased swimming of microalgae toward a light source that produces a focusing of the microswimmer suspension. We combine experiments that investigate the stationary state of this phenomenon as well as the transition regime with \eb{analytical and numerical} modeling. We show that the experimentally observed scalings \eb{on the width of the focalized region and the establishment length as a function of the flow velocity} are well described by a simple theoretical model. 
\end{abstract}

\maketitle


Microorganisms such as phytoplankton have been shown very often to present some patchiness in their spatial distribution \cite{patchiness, Bees1998}.
\eb{These} spatial inhomogeneities occur either at large scales as in the ocean or at smaller scales in lakes or ponds for instance. Although this patchiness might in some case be due to inhomogeneities of the ecosystem -- gradients in temperature \eb{or} nutrients, gravity, etc. -- the coupling between flow and plankton's ability to swim is suspected to play a major role in the spatial localization of plankton. Indeed, patchiness has been shown to be more important in the case of swimming as compared to non-swimming plankton \cite{durham2013turbulence}.

More generally, swimming plankton represents one among many systems that form what is called active matter \cite{marchetti2013}, i.e., systems composed of units able to self-propel autonomously. These can be molecular motors, bacteria, algae, fish to only mention some living systems.
Active matter attracts a lot of attention from scientists as it represents a new "state of (non-equilibrium) matter" to be investigated. In particular, an interesting feature to understand is how the coupling between flow and motility occurs \cite{lauga2009hydrodynamics, barry2015shear}. This question has been addressed for instance in the case of gyrotactic algae in \eb{a} flow \cite{Kessler1985a,boffetta}.
Because of the eccentric location of their chloroplast, these algae are subject to a torque \eb{induced by gravity} and consequently tend to swim upward. Gyrotactic algae have been shown to concentrate in downwelling regions of turbulent flows \cite{boffetta}; this is a direct consequence of the migration of gyrotactic algae towards low vorticity regions as shown by Kessler \cite{Kessler1985a}.

Here, we investigate an analogous phenomenon that appears when flow vorticity is coupled to phototaxis --a biased swimming of algae toward a light source.
We propose a detailed quantitative description of the experimental dynamics and stationary regimes of what we call photofocusing \cite{garcia2013} coupled to an analytical and numerical treatment of the problem \eb{in a spirit} similar to \cite{zottl, delillo2014}.

\subsection{Experimental part}
\begin{figure}
\begin{center}
\includegraphics[width=0.8\columnwidth]{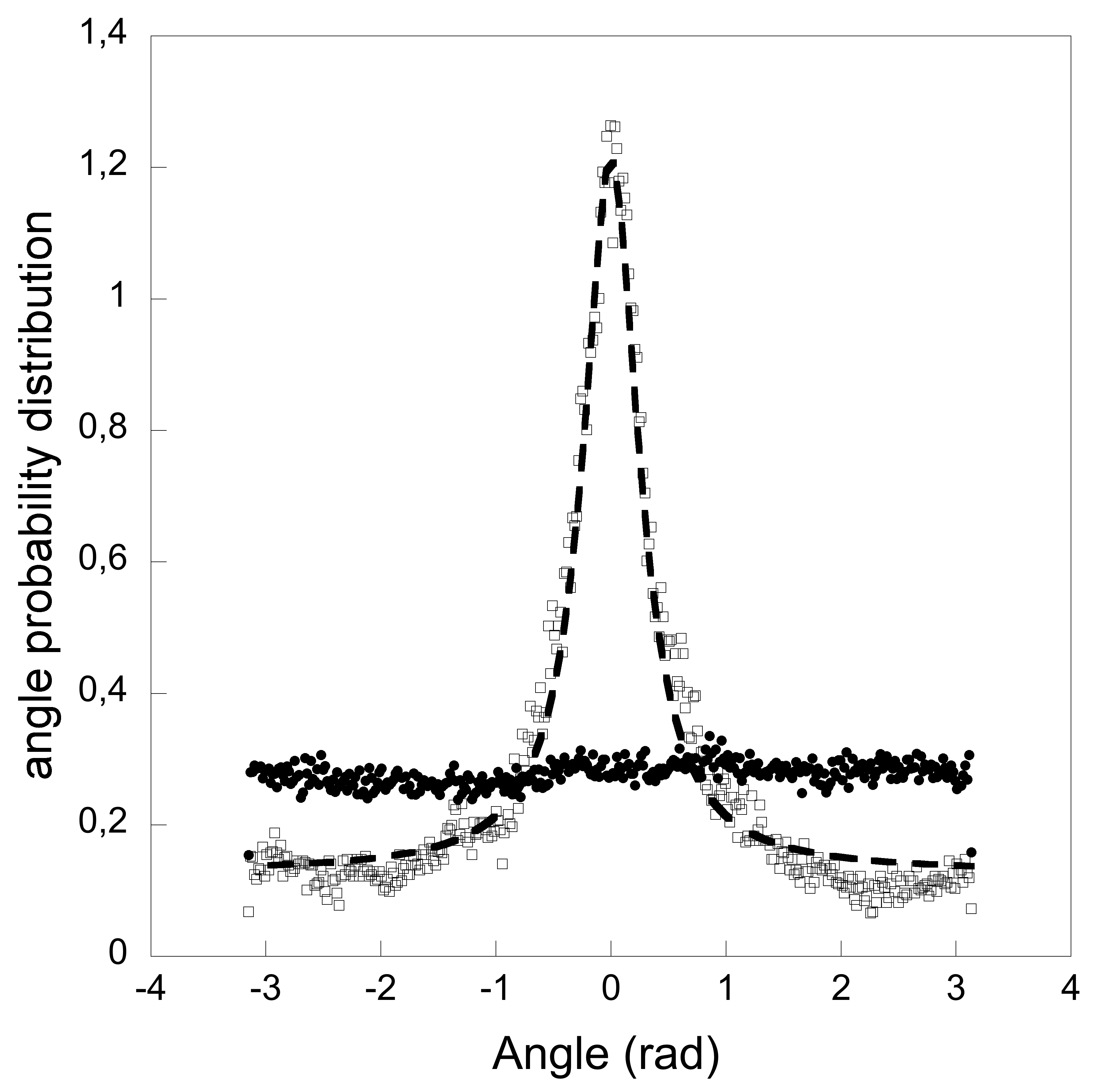}
\caption{\label{fotoangle} Probability distribution of angles $\psi(\theta-\pi)$  in absence of light source (filled symbols) and when a light is switched on in the direction $\theta = \pi$ (open symbols); the cells are swimming in a fluid at rest. A truncated Lorentzian of width $1.2$ is found to adjust the experimental points.
}
\end{center}
\end{figure}

 \begin{figure}
 \begin{tabular}{ll}
 a \\
\includegraphics[width=0.8\columnwidth]{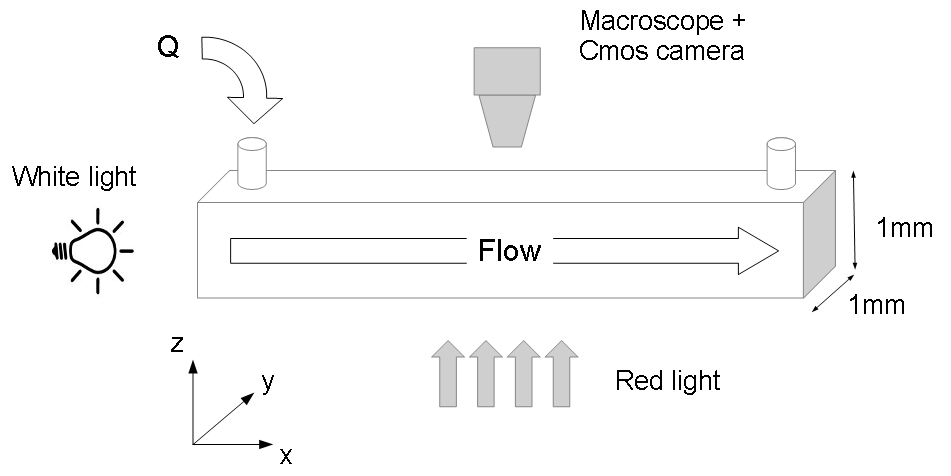}\\
 b  \\
\includegraphics[width=0.8\columnwidth]{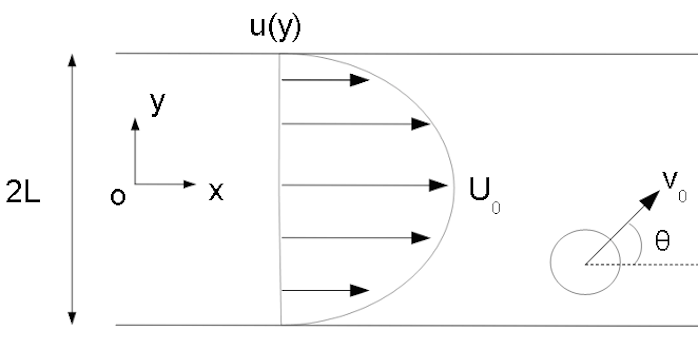}\\
 \end{tabular}
\caption{\label{setup} Experimental setup and coordinates system.}
\end{figure}

The green microalga \textit{Chlamydomonas Reinhardtii} is a biflagellated photosynthetic cell of 10 micrometer diameter \cite{sourcebook}. Cells are grown under a 14h/10h light/dark cycle at 22$^{o} $C. Cells are harvested in the middle of the exponential growth phase. This microalga propels itself in a break-stroke type swimming using its two front flagella \cite{PolinGoldstein2009}. This way of swimming has been shown to be well characterized by a persistent random walk \cite{garcia2011,PolinGoldstein2009} in absence of tropism. The typical time of persistence is of the order of a few seconds.
When subjected to a light stimulus (green wavelength, i.e., around 510 nm), cells tend to swim toward the light source and perform a quasi ballistic motion.

\eb{Let us} first quantify the useful characteristics of the phototaxism mechanism of the cells.
Whereas in absence of light bias, microalgae perform a persistent random walk, they adopt a ballistic motion toward the direction of a light source. Their response time is related to the mean reorientation time in the random walk of the cells; indeed, these reorientations allow the cells to scan the space for the presence of a light source. This time can be quantified by measuring the time correlation function of direction over a whole population of cells as a function of the time \cite{garcia2011} and is found to be of the order of $3$s. 

Once microswimmers swim toward the light, trajectories still show a dispersion in their orientation. Figure~\ref{fotoangle} shows the probability distribution of angles $\psi(\theta-\pi)$  in absence of light source and when a light is switched on in the direction $\theta = \pi$ in a fluid at rest.

The cells phototaxism is then coupled to a Poiseuille flow. A squared section channel is made of PDMS ($1\times 1\times 50$ mm$^{3}$). A white led is used as a light source placed upstream (fig.~\ref{setup}). A syringe pump imposes a flow with flow rates Q ranging from 0.5 to 3 mm$^{3}.$s$^{-1}$.
Observation is made with a macroscope (Leica APOZ16) coupled to a CMOS camera. A low magnification objective is used that provides a field of view of size $1.5$ mm $\times 20$ mm; this insures the observation of a stationary state (contrary to our previous study \cite{garcia2013}). The flowing channel is enclosed in an occulting box with two red filtered windows for visualization. This is to avoid any parasite light that would trigger phototaxis.

\begin{figure}
 \begin{tabular}{ll}
 a & b \\
\includegraphics[width=0.5\columnwidth]{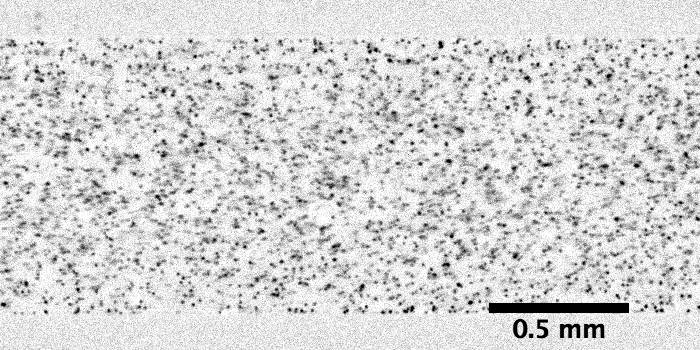} &\includegraphics[width=0.5\columnwidth]{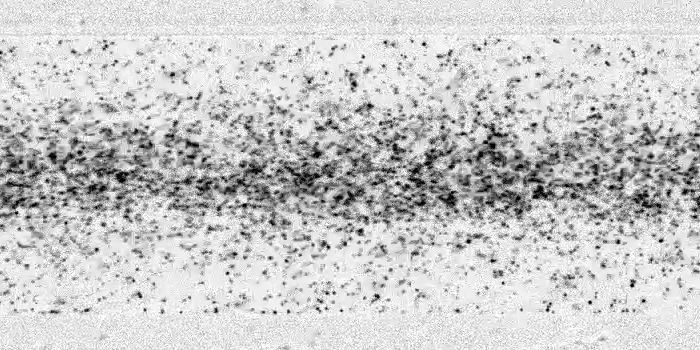}\\
 \end{tabular}
\caption{\label{photo} Examples of typical experiments. In absence of light (a), microswimmers are dispersed uniformly along y direction; in presence of light upstream (b), the cells are focalized in the central region of the channel. Note that here only zoomed pictured are shown for the sake of clarity but the experiments have explored a much wider field of view: 1.5 mm x 20 mm.}
\end{figure}

\begin{figure}
\begin{center}
\includegraphics[width=\columnwidth]{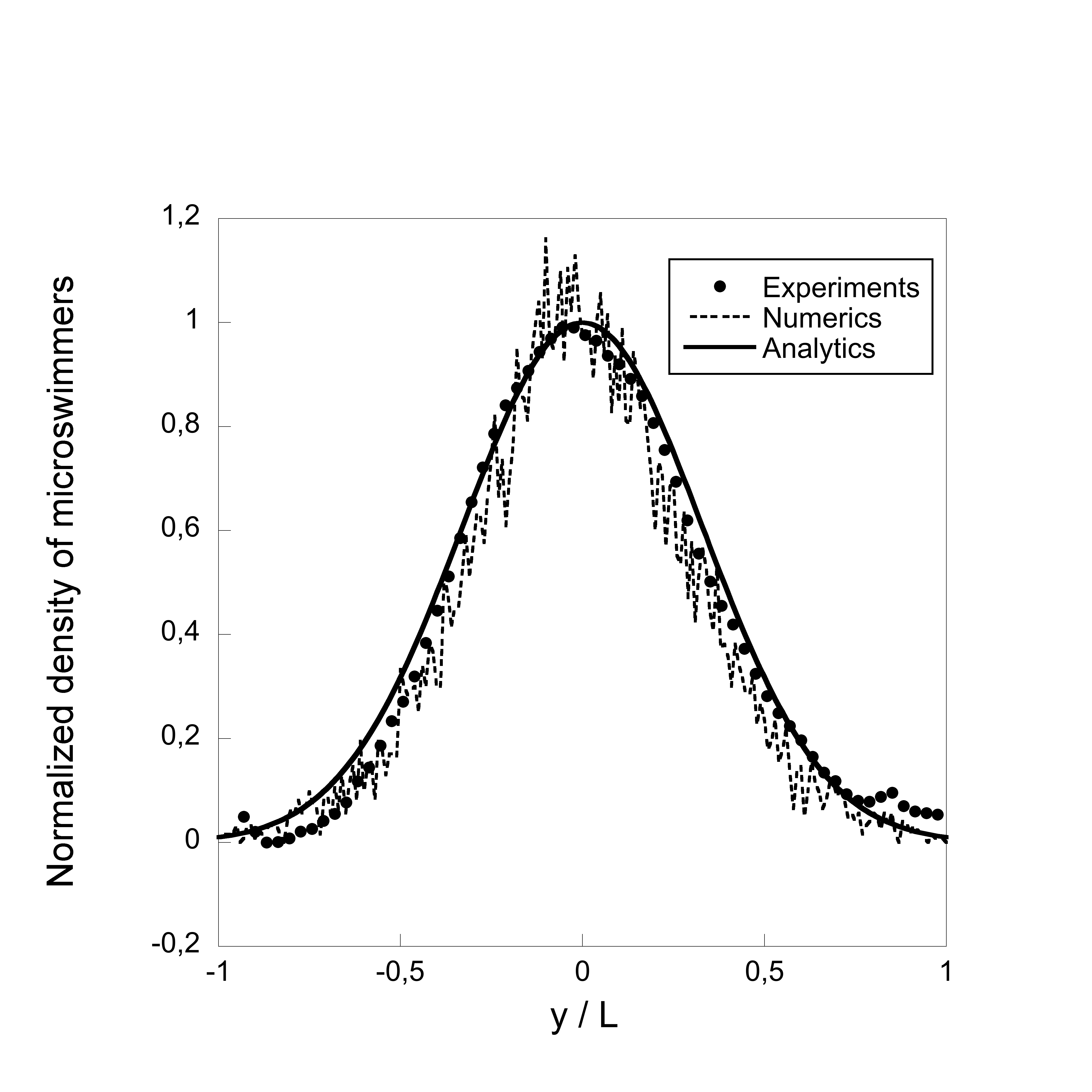}
\caption{\label{Q70distrib} Dots: Experimental density distribution of cells deduced from the logarithm of the grey level intensity of the images (flow rate of 70 $\mu$L/min and light source upstream, only half of data points are shown for clarity). Dashed line: Numerical density distribution. Plain line: Analytical expression of the density distribution (eq.~\ref{eq:rho-yGauss}).}
\end{center}
\end{figure}

In the stationary state, images of the focused cells (fig.~\ref{photo}) are taken and analyzed to deduce the concentration profile along $y$ direction. For this, Beer-Lambert law is used to convert \eb{grey} levels of intensity into concentration values. Volume fractions in the experiments are below $0.4\%$.  The cell density distribution along $y$ direction is then measured as a function of flow rate and in presence of light upstream. An example of  distribution is shown in \eb{fig.}~\ref{Q70distrib} and compared to the analytical and numerical distributions (the models are described in the following sections). We measure the exponential length $L_{x}$ over which a stationary density profile is reached. This length is plotted as a function of the flow velocity in fig.~\ref{lx}. The square root $d_{\infty}$  of the variance of stationary experimental distributions are then extracted and plotted as a function of flow mean velocity (fig.~\ref{fig:d}); the experimental results are shown to be well described by a continuum model presented in the following.

\subsection{Analytical modeling}

\subsubsection{\eb{Dynamics of individual particles}}

We consider a two-dimensional Poiseuille flow, seeded with point-like microswimmers which tend to reorient their swimming direction towards the light source, situated up-stream, at randomly chosen times. We denote as $x$ the direction of the flow, $y$ the direction transverse to the flow, and $2L$ the width of the channel ($-L\le y\le L$). The velocity profile of the flow is given by
\begin{equation}
u(y) = U_0 \left( 1-\frac{y^2}{L^2} \right)
\end{equation}
and the strain rate is:
\begin{equation}
\label{eq:shearrate}
\dot\gamma = \frac{du}{dy} = -\frac{2U_0}{L^2} y.
\end{equation}

Each swimmer moves at a constant speed $v_0$ in its own swimming direction, characterized by an angle $\theta$ (fig.~\ref{setup}) which evolves in time due to two different contributions. \eb{First}, due to the vorticity of the flow, the swimmer rotates at an angular velocity
\begin{equation}
\omega = -\frac{\eta \dot\gamma}{2} = \frac{\eta U_0}{L^2}\, y
\end{equation}
where $\eta$ is a partial drive coefficient ($0<\eta\le 1$) resulting from the fact that the microswimmers partially resist against the vorticity of the flow \cite{Rafai2010}. Secondly, each swimmer reorients, at randomly chosen times, its direction of motion close to the direction of the light source (fig.~\ref{fotoangle}).
More precisely, a new angle $\theta$ is chosen from a probability distribution $\psi(\theta-\pi)$, where $\psi(\theta')$ is a distribution centered around $\theta'=0$, and where $\theta=\pi$ corresponds to the exact direction of the light source (fig.~\ref{fotoangle}). Reorientation times occur randomly with rate $\alpha$, meaning that the time duration $\tau$ between two successive reorientations is exponentially distributed, $p(\tau)=\alpha\, e^{-\alpha \tau}$.

\subsubsection{\eb{Statistical description}}

We assume that the concentration of swimmers is small enough so that hydrodynamic interactions between swimmers can be neglected.
The key quantity to describe the statistics of swimmers is the probability
$f({\bf r},\theta,t)$ for a swimmer to be at position ${\bf r}=(x,y)$, with a direction $\theta$, at time $t$. The evolution equation for $f$ reads
\begin{equation} \label{eq:evol:f}
\partial_t f + \big( u(y) \mathbf{e}_x + v_0 \mathbf{e}(\theta) \big) \cdot \nabla f + \frac{\eta U_0 y}{L^2}\, \partial_{\theta} f
= - \alpha f + \alpha\rho\, \psi(\theta-\pi)
\end{equation}
where $\rho$ is the \eb{concentration} of swimmers,
\begin{equation}
\rho({\bf r},t) = \int_0^{2\pi} f({\bf r},\theta,t) \, d\theta
\end{equation}
and $\mathbf{e}_x$ is the unit vector along the $x$-direction (direction of the flow), while $\mathbf{e}(\theta)$ is the unit vector in the direction $\theta$, with $\mathbf{e}(\theta=0)=\mathbf{e}_x$. Eq.~(\ref{eq:evol:f}) contains three different contributions to the evolution of the probability $f({\bf r},\theta,t)$. The second term on the l.h.s.~of Eq.~(\ref{eq:evol:f}) corresponds to the advection of particles under the combined effect of self-propulsion and Poiseuille flow. The third term on the l.h.s.~describes the rotation of the swimmers induced by the shear flow. Finally, the r.h.s.~of Eq.~(\ref{eq:evol:f}) describes the reorientation dynamics, through which the current angle $\theta$ is instantaneously changed into a new angle drawn from the distribution $\psi(\theta-\pi)$.
It is convenient to define dimensionless variables by taking the channel half-width $L$ as the unit of length, and the inverse of the reorientation rate $\alpha$ as the unit of time:
\begin{equation}
\tilde{x}=\frac{x}{L} \;, \qquad \tilde{y}=\frac{y}{L} \;, \qquad \tilde{t}=\alpha t  \;, \qquad \tilde{f}=L^2 f \;.
\end{equation}
In the following, we drop the tildes to lighten notations.
The dimensionless evolution equation for $f$ is then given by
\begin{equation} \label{eq:f-adim}
\partial_t f + \Big( \frac{b}{\eta} (1-y^2) {\bf e}_x +\beta \, {\bf e}(\theta)\Big) \cdot \nabla f + b y\, \partial_{\theta} f
= - f + \rho\, \psi(\theta-\pi)
\end{equation}
with $-1 \le y \le 1$ and
\begin{equation}
\beta = \frac{v_0}{\alpha L}  \;, \qquad
b = \frac{\eta U_0}{\alpha L} \;.
\end{equation}
In the experiment, one has $v_0 \approx 10^{-1} $mm.s$^{-1}$, $\alpha \approx 0.33$ s$^{-1}$, $2L=1$ mm and a shear rate of a few s$^{-1}$. With these values, one finds $\beta \approx 0.6$ and the parameter $b$, which measures the ratio between the shear rate and the reorientation rate, is of the order of $5$.

\begin{figure}[htbp]
\begin{center}
\includegraphics[width=\columnwidth]{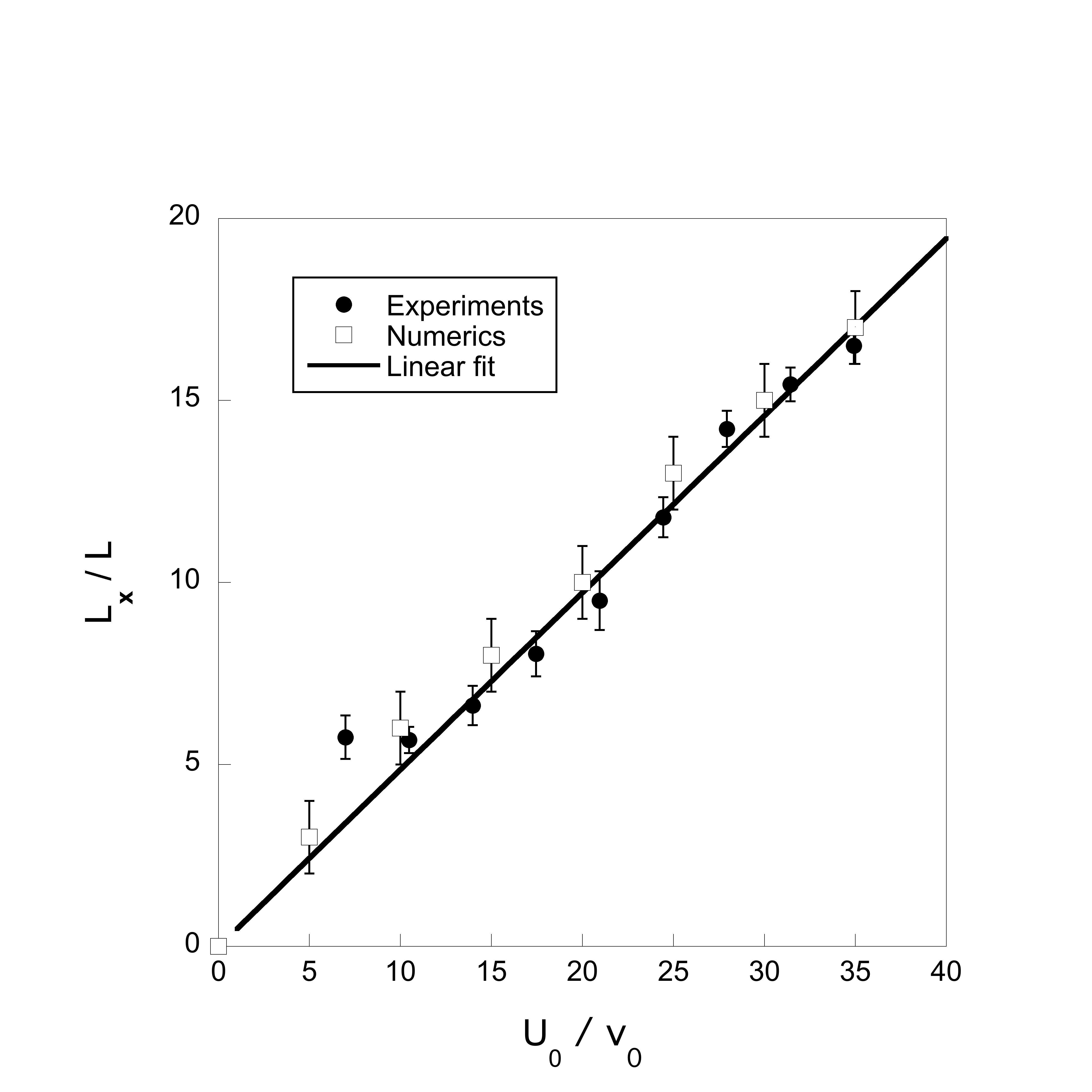}
\caption{Normalized focusing length scale $L_{x}$  as a function of the normalized flow speed $U_{0}/v_{0}$. Experimental data (dots) and numerical data (squares). \eb{Data are obtained by varying $U_0$.} The fit of the experimental data represents a linear scaling as suggested in the text $L_{x} \approx U_{0}/\alpha$.}
\label{lx}
\end{center}
\end{figure}

Considering the coefficient $b/\eta$ of the $x$-derivative in Eq.~(\ref{eq:f-adim}) suggests that the density profile relaxes, from the entrance of the channel, to its $x$-independent shape over a typical length scale $L_x \approx (b/\eta) L=U_0/\alpha$. Figure~\ref{lx} shows the experimentally determined rescaled focusing length in the x-direction and the numerical values as a function of the rescaled flow speed. Both are well described by a linear law with a prefactor of $0.5$ which is consistent with the expected value of $v_{0}/\alpha L = 0.6$.

\subsubsection{\eb{Analytical determination of the concentration profile}}

We now wish to determine an analytical approximation of the stationary concentration profile far from the entrance of the channel, where the profile becomes invariant along the direction of the channel.
Starting from Eq.~(\ref{eq:evol:f}), we expand the distribution
$f({\bf r},\theta)$ in angular Fourier modes, yielding a hierarchy of coupled equations. Using a simple closure relation, this hierarchy can be truncated at second (nematic) order, allowing for a simple Gaussian solution of the density profile given by
\begin{equation} \label{eq:rho-yGauss}
\rho(y) = \rho_{\rm max} \, e^{-y^2/2\sigma^2}
\end{equation}
with
\begin{equation}
\sigma = \sqrt{\frac{(1-\psi_2)}{2\eta\psi_1}\, \frac{v_0}{U_0}}
\label{eq:sigma-Gauss}
\end{equation}
\eb{The coefficients} $\psi_k$ are the \eb{Fourier modes} of the distribution $\psi(\theta)$ (fig.~\ref{fotoangle}).
Technical aspects of the calculations are reported in Appendix~\ref{sec-AppA}.
The expression (\ref{eq:sigma-Gauss}) of the width $\sigma$ confirms the scaling in $\sqrt{v_0/U_0}$ observed in experimental data.

\begin{figure}
\begin{center}
\includegraphics[width=\columnwidth]{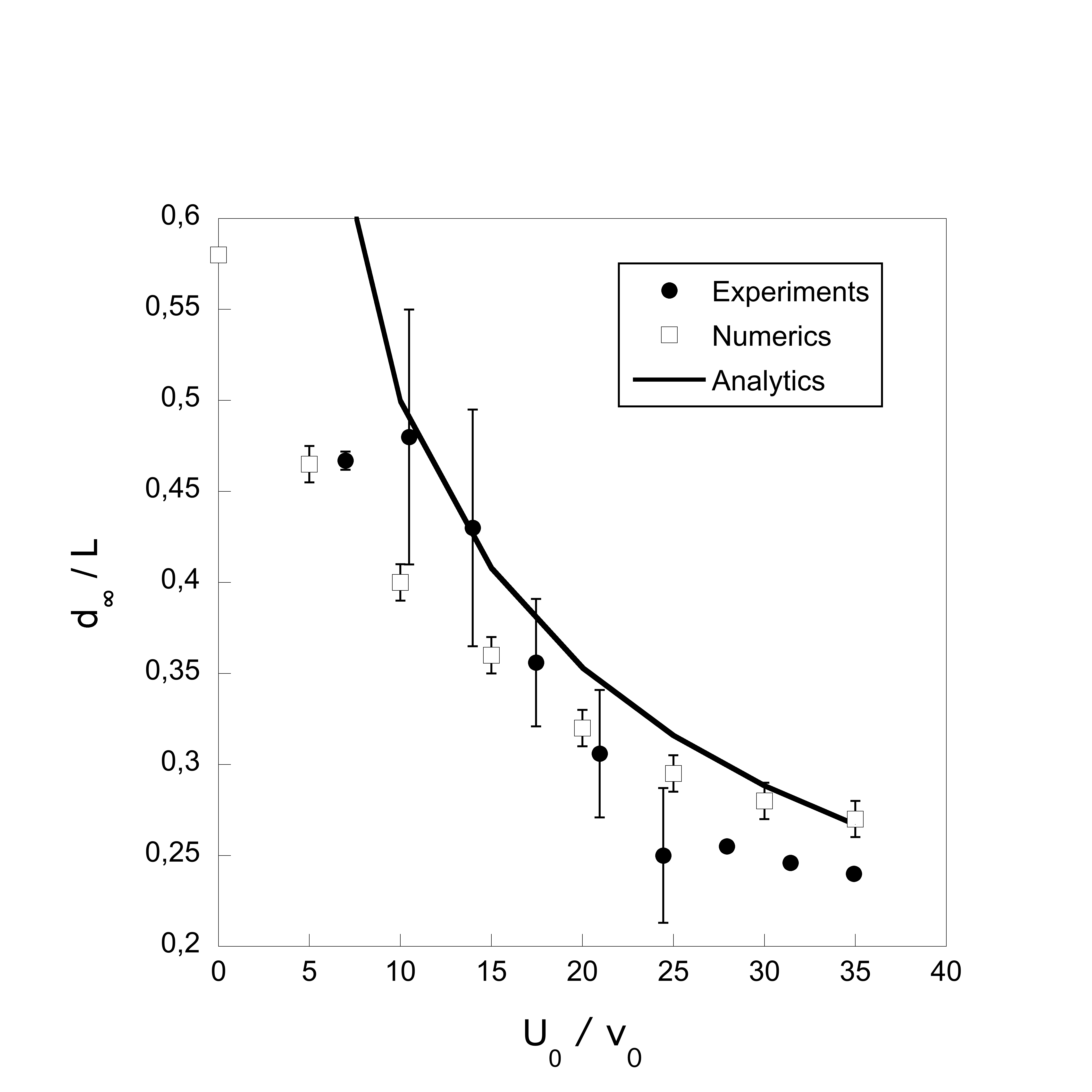}
\caption{\label{fig:d} Half width of the cell density profile rescaled by the channel half width as a function of the rescaled flow speed. Filled symbols represent the experimental data points. Squared symbols represent the numeral data points and the plain line is calculated from the variance of the analytical expression of the density distribution (Eq.\ref{eq:sigma-Gauss}).}
\end{center}
\end{figure}

\subsection{Numerical simulations}

We consider a swimmer moving with a velocity \eb{$\mathbf{v_{0}}=v_0 \mathbf{e}(\theta)$} in
a Poiseuille flow along the $x-$direction. 
\eb{Its velocity $\mathbf{v}$ in the lab frame is given by}
\begin{equation}
\left\{ \begin{array}{c}
v_{x}=v_{0}\cos\theta+U_{0}\left[1-\left(y/L\right)^{2}\right]\\
\\
v_{y}=v_{0}\sin\theta
\end{array}\right.,\label{eq: vitesse}
\end{equation}
where $\theta$ is, \eb{as previously,} the angle between the swimmer and the $x-$axis
(if $0<\vert\theta\vert<\pi/2$, the swimmer is oriented downstream, \eb{while} if $\pi/2<\vert\theta\vert<\pi$,
the swimmer is oriented upstream). The swimmer is constantly rotated
by the vorticity $\dot{\gamma}$ (see Eq.~\ref{eq:shearrate}) of the external flow:
\begin{equation}
\theta(t)=\int_{0}^{t} \eb{\dot{\gamma}(y(t'))\, dt' \;.}
\end{equation}
 The light being \eb{situated} upstream, the angle is set to $\pi$ at a frequency
$\alpha$, where $\alpha^{-1}$ is the persistence time which corresponds
to a ballistic motion of a cell in a fluid at rest. The parameter
$\eta$ corresponds to the fact that a cell resists the flow rotation
as observed experimentally (\textit{i.e.} $\eta<1)$ \cite{Rafai2010}. Here, we find $\eta = 0.25$ \sr{as a single adjustable parameter when fitting both $d_{\infty}$ and $L_{x}$.} In our simulations,
we calculate the trajectories of $10\,000$ swimmers by integrating
Eq.(\ref{eq: vitesse}), $\mathbf{r}(t)= \int_{0}^{t} \eb{\mathbf{v}(t')dt'}$
with different initial conditions: $-L<y<L$ and $x=0$. Then, we
calculate the distribution of swimmers accross the channel for different
$x-$values $\rho(x,y)$ and evaluate the width of this distribution:
$d(x)/L=\left[\int_{-L}^{L}\left(y/L\right)^{2}\rho(x,y)dy\right]^{1/2}$.
From $d(x)/L$, we evaluate the length $L_{x}$ (fig.~\ref{lx}) on which
the focusing reaches its stationnary values $d_{\infty}$ (fig.~\ref{fig:d}).

\subsection{Discussion \eb{and conclusion}}

In this paper, we show that the photofocusing experimentally observed can be well described both by numerical and analytical models  which neglect hydrodynamics interactions (a reasonable assumption for such dilute suspensions).  The experiments described in this paper allowed \eb{us} to investigate both the transition regime of the photofocusing phenomenon as well as the stationary distributions of cells in the channel. An analytical continuous model is shown to satisfactorily describe the \eb{experimental data} and provides us with useful scaling laws. Moreover, numerical simulations \eb{confirm} these findings.
Interestingly, the \eb{consistency between experimental data and modeling backs} up a hypothesis that we previously emitted concerning the resistance of the cells to vorticity and that is a key ingredient to understand the peculiar rheological behaviour of Chlamydomonas Reinhardtii suspensions \cite{Rafai2010}.

\acknowledgements{The authors thank A. Chardac for measuring the probability distribution of angles in \eb{the} presence of light. This work has been partially supported by the LabEx Tec 21 (Investissements d'Avenir - grant agreement n° ANR-11-LABX-0030)}


\appendix

\section{Analytical derivation of the density profile}
\label{sec-AppA}

We provide in this Appendix the technical aspects of the analytical derivation of the velocity profile given in Eq.~(\ref{eq:rho-yGauss}).
Let us define, for integer $k$, the angular Fourier mode as
\begin{equation}
f_k({\bf r},t) = \int_0^{2\pi} f({\bf r},\theta,t) \, e^{ik\theta} \, d\theta
\end{equation}
Note that $f_{-k}=f_k^*$, where $f_k^*$ denotes the complex conjugate of $f_k$, and that $f_0=\rho$.
Expanding Eq.~(\ref{eq:evol:f}) in Fourier modes then leads to
\begin{eqnarray} \nonumber
\partial_t f_k &+& U_0 \left( 1-\frac{y^2}{L^2} \right) \partial_x f_k
+ \frac{v_0}{2} ( \hat\triangledown f_{k-1} + \hat\triangledown^* f_{k+1})
\\
&& \qquad - ik\eta U_0 y f_k  = -\alpha f_k + (-1)^k \alpha \psi_k\, \rho
\label{eq:f-Fourier}
\end{eqnarray}
with $\psi_k$ the corresponding Fourier mode of the distribution $\psi(\theta)$,
\begin{equation}
\psi_k = \int_0^{2\pi} \psi(\theta) \, \cos (k\theta) \, d\theta \, ,
\end{equation}
(note that the symmetry $\psi(-\theta)=\psi(\theta)$ has been taken into account). To shorten notations, we have introduced in Eq.~(\ref{eq:f-Fourier}) the complex differential operators
\begin{equation}
\hat\triangledown = \partial_x + i\partial_y \;, \quad
\hat\triangledown^* = \partial_x - i\partial_y \,.
\end{equation}
For $k=0$, Eq.~(\ref{eq:f-Fourier}) reduces to the continuity equation
\begin{equation}
\partial_t \rho + U_0 \left( 1-\frac{y^2}{L^2} \right) \partial_x \rho + v_0 {\rm Re}(\hat\triangledown^* f_1) = 0 \,.
\label{eq-rho}
\end{equation}
Note that $v_0 f_1$ is the exact analog, using the natural mapping between complex numbers and two-dimensional vectors, of the usual mass flux $\rho \bar{\bf v}$, where $\bar{\bf v}$ is the local collective velocity of swimmers. Hence Eq.~(\ref{eq-rho}) is equivalent to a standard continuity equation of the form
\begin{equation}
\partial_t \rho + \nabla \cdot (\rho \bar{\bf v}) =0,
\end{equation}
with the mass current $\rho \bar{\bf v}$ given by
\begin{equation} \label{eq:continuity}
\rho \bar{\bf v} = \Big[ U_0 \left( 1-\frac{y^2}{L^2} \right) \rho
+ v_0 {\rm Re}(f_1)\Big] \mathbf{e}_x + v_0 {\rm Im}(f_1)\, \mathbf{e}_y
\end{equation}
where ${\rm Re}(z)$ is the real part of the complex number $z$, and
${\rm Im}(z)$ its imaginary part.

We now search for the stationary density profile $\rho(y)$, assumed to be invariant along the $x$-direction, that is the direction of the flow.
Similarly, $f_1$ depends only on $y$.
Under these assumptions, we find that in the stationary state,
$f_1$ satisties $\partial_y \, {\rm Im} f_1(y)=0$.
Hence ${\rm Im} f_1(y)$ is a constant, independent of $y$.
Given that ${\rm Im} f_1(y)$ is proportional to the flux of swimmers in the $y$-direction [see Eq.~(\ref{eq:continuity})], ${\rm Im} f_1$ has to be zero at the walls of the channel, implying ${\rm Im} f_1=0$ for all $y$.
The density profile is thus obtained by solving ${\rm Im} f_1=0$, which requires to be able to express $f_1$ as a function of $\rho$ and its derivatives.
This can be done be taking into account the evolution equation of the mode $f_1$, namely Eq.~(\ref{eq:f-Fourier}) for $k=1$, which reads
\begin{equation}
\frac{iv_0}{2} ( \partial_y \rho -\partial_y f_2) - i\eta U_0 y f_1 = -\alpha f_1 - \alpha \psi_1\, \rho \;.
\label{eq-f1} 
\end{equation}
From this equation, one can express $f_1$ as a function of $\partial_y \rho$ and $\partial_y f_2$,
\begin{equation} \label{eq:f1-dyrho-dyf2}
f_1 = \frac{1}{\alpha-i\eta U_0 y} \, \left[ -\alpha \psi_1 \rho - \frac{iv_0}{2} ( \partial_y \rho - \partial_y f_2) \right] \;.
\end{equation}
In order to close the equation, we need to express $f_2$ as a function of $\rho$ and $f_1$ (and possibly their derivatives).
Writing Eq.~(\ref{eq:f-Fourier}) for $k=2$, one finds
\begin{equation}
\frac{iv_0}{2} ( \partial_y f_1 -\partial_y f_3) - 2i\eta U_0 y f_2 = -\alpha f_2 + \alpha \psi_2\, \rho \;.
\label{eq-f2}
\end{equation}
This equation also involves the higher order mode $f_3$. To obtain a simple expression for $f_2$, we thus need to make an approximation in order to close the set of equations.
We first note that in the absence of flow ($U_0=0$) and self-propulsion ($v_0=0$), the distribution $f(\theta)$ simply relaxes to ${\bar f}(\theta)=\rho \psi(\theta)$, yielding for the Fourier modes ${\bar f}_k = \rho \psi_k$.
The flow and self-propulsion can then be considered as driving mechanisms that perturb this distribution ${\bar f}(\theta)$. A simple closure relation is then to neglect the effect of these perturbations on $f_2$ (while, of course, keeping them on $f_1$).
We thus simply assume in the following that $f_2=\rho \psi_2$.
Under this approximation, the equation ${\rm Im} f_1=0$ can be rewritten as,
using Eq.~(\ref{eq:f1-dyrho-dyf2}),
\begin{equation}
\eta U_0 \psi_1 y \rho + \frac{v_0}{2} (1-\psi_2) \partial_y \rho =0 \;.
\end{equation}
After integration, we obtain for the density profile
\begin{equation}
\rho(y) = \rho_{\rm max} \, \exp\left( -\frac{\eta U_0 \psi_1}{v_0(1-\psi_2)}\, y^2 \right)
\end{equation}
(where $\rho_{\rm max}$ is a constant), as given in Eq.~(\ref{eq:rho-yGauss}).

\bibliographystyle{apsrev}
\bibliography{chlamybib2}

\end{document}